\documentclass[aps,reprint,superscriptaddress]{revtex4-1}
\usepackage{amsmath}
\usepackage{graphicx}
\usepackage{textgreek}
\usepackage{fixltx2e}
\usepackage{hyperref}
\usepackage{nicefrac}
\usepackage[utf8]{inputenc}
\usepackage{color}

\draft
\begin{document}
\title{Robust measurement of supernova $\nu_e$ spectra with future neutrino detectors}
\author{Alex Nikrant}
\email{anik417@vt.edu}
\affiliation{Department of Physics, Virginia Tech, Blacksburg, Virginia 24061, USA}
\author{Ranjan Laha}
\email{ranjalah@uni-mainz.de}
\affiliation{KIPAC, Department of Physics, Stanford University, Stanford, California 94305, USA }
\affiliation{SLAC National Accelerator Laboratory, Menlo Park, California 94025, USA}
\affiliation{PRISMA Cluster of Excellence and Mainz Institute for Theoretical Physics, Johannes Gutenberg-Universitat Mainz, 55099 Mainz, Germany}
\author{Shunsaku Horiuchi}
\email{horiuchi@vt.edu}
\affiliation{Department of Physics, Virginia Tech, Blacksburg, Virginia 24061, USA}
\affiliation{Center for Neutrino Physics, Department of Physics, Virginia Tech, Blacksburg, Virginia 24061, USA}
\date{\today}

\begin{abstract}
Measuring precise all-flavor neutrino information from a supernova is crucial for understanding the core-collapse process as well as neutrino properties. We apply a chi-squared analysis for different detector setups to explore determination of $\nu_{e}$ spectral parameters.  Using a long-term two-dimensional core-collapse simulation with three time-varying spectral parameters, we generate mock data to examine the capabilities of the current Super-Kamiokande detector and compare the relative improvements that gadolinium, Hyper-Kamiokande, and DUNE would have. We show that in a realistic three spectral parameter framework, the addition of gadolinium to Super-Kamiokande allows for a qualitative improvement in $\nu_e$ determination. Efficient neutron tagging will allow Hyper-Kamiokande to constrain spectral information more strongly in both the accretion and cooling phases. Overall, significant improvements will be made by Hyper-Kamiokande and DUNE, allowing for much more precise determination of $\nu_e$ spectral parameters.
\end{abstract}

\maketitle

\section{Introduction}

Supernovae are among the most spectacular events in the Universe, and can briefly outshine their entire host galaxies.  In a core-collapse supernova (CCSN), the core of a massive star reaches a critical mass and collapses from thousands of kilometers to tens of kilometers in a fraction of a second, releasing an immense amount of gravitational binding energy ($\sim$ 3 $\times$ 10$^{53}$ erg).  A full understanding of the core-collapse phenomenon requires different areas of physics to come together in an intricate interplay.  Currently, we do not fully understand how the collapse reverts to form an explosion, and neutrinos can offer a unique window into this process\,\cite{Janka:2006fh,Woosley:2006ie,Burrows:2006ci,Mezzacappa:2005ju,Janka:2012wk}.  

During a CCSN (the death of stars having a zero age main sequence mass $\gtrsim$ 8 M$_\odot$), $\sim$ 99\% of the gravitational binding energy released is emitted in the form of neutrinos and antineutrinos.  Neutrinos are weakly interacting and they make an unimpeded journey from the collapsed star to terrestrial detectors.  They can tell us information about the core of the CCSN, which is otherwise unreachable using photon observations since it is hidden by the stellar envelope.  This makes neutrinos a unique and crucial part in understanding CCSNe\,\cite{Horiuchi:2008jz,Lund:2010kh,Scholberg:2012id,Kistler:2012as,Tamborra:2013laa,Cherry:2013mv,Tamborra:2014aua,Vissani:2014doa,Tamborra:2014hga,Patton:2014lza,Tian:2016hec,Horiuchi:2017sku,Wright:2017jwl,GalloRosso:2017hbp,Kneller:2017lqg,Horiuchi:2017qlw,Horiuchi:2017qja}.  The flux of neutrinos and antineutrinos integrated over all energies from a Galactic supernova (assumed to be at a distance of 10 kpc from Earth) is $\mathcal{O}$($10^{11}$ cm$^{-2}$ s$^{-1}$).  With current detectors, there will be a reasonable number of neutrino event detections, $\sim \mathcal{O}$(10-100) for each flavor of neutrino, with the exception of $\bar{\nu}_{e}$, which yields $\sim \mathcal{O}$(10,000), mostly due to the large inverse beta cross section at CCSN energies.  

So far, we have detected neutrinos only from one supernova, SN 1987A.  Between Kamiokande-II\,\cite{Hirata:1987hu} and the Irvine-Michigan-Brookhaven water Cherenkov detector \cite{PhysRevLett.58.1494}, some two dozen events were observed.  Despite the low number of events, much has been learned about CCSN neutrino emission from this detection\,\cite{PhysRevLett.58.1906,FRIEMAN1988115,kahana_1988,PhysRevLett.58.2722,Costantini:2004ry,Mirizzi:2005tg,Costantini:2006xd,Pagliaroli:2008ur,Ianni:2009bd,Vissani:2014doa}.  Today, detection prospects have dramatically improved with the construction of Super-Kamiokande (Super-K), and further improvements are expected with future detectors such as JUNO\,\cite{An:2015jdp}, DUNE\,\cite{Kudryavtsev:2016ybl,Acciarri:2016ooe}, and Hyper-Kamiokande (Hyper-K)\,\cite{Abe:2011ts,Hyper-Kamiokande:2016dsw}.

One of the active areas of research with CCSN neutrinos is understanding flavor mixing, especially neutrino collective oscillations\,\cite{Dighe:1999bi,Dighe:2007ks,Duan:2010bg,Cherry:2011fm,Cherry:2011fn,Tamborra:2013laa,Raffelt:2013isa,Mirizzi:2015eza,Dasgupta:2015iia,Chakraborty:2016lct,Izaguirre:2016gsx,Akhmedov:2016gzx,Armstrong:2016mnt,Capozzi:2016oyk,Johns:2016wjd,Chakraborty:2016yeg,Tamborra:2017ubu,Capozzi:2017gqd,Sasaki:2017jry,Akhmedov:2017mcc,Dasgupta:2017oko}. CCSN neutrinos can provide crucial information needed to refine our understanding of collective oscillations but only if we can get all-flavor information from detectors.  We can also learn about the mass hierarchy and nonstandard neutrino interactions using precise flavor information of CCSN neutrinos\,\cite{Mirizzi:2015eza,Lai:2016yvu,Stapleford:2016jgz,Das:2017iuj,Dighe:2017sur}.  Thus, it is important to be able to precisely measure parameters of each flavor of CCSN neutrinos.  Water Cherenkov detectors, like Super-K, have excellent capabilities of detecting $\bar{\nu}_{e}$ via the inverse beta interaction. High event numbers for this reaction are expected, so analyses of $\bar{\nu}_{e}$ will be relatively precise \cite{Scholberg:2012id}.  Nonelectron flavor neutrinos, $\nu_{\mu}$, $\nu_{\tau}$, and $\bar{\nu}_\mu$, $\bar{\nu}_\tau$ (referred to as $\nu_{x}$ and $\bar{\nu}_x$, due to their similar behavior in the context of CCSNe), have only neutral current interactions and thus yield fewer event numbers.  These can be detected in large liquid scintillator detectors with low detection energy thresholds\,\cite{Beacom:2002hs,Dasgupta:2011wg,Lujan-Peschard:2014lta}.  Neutral current reactions in water can also be used to detect $\nu_x$ and $\bar{\nu}_x$\,\cite{Langanke:1995he}.

We do not currently have one particular setup that will detect $\nu_{e}$ with great efficiency.  Among current detectors, the interaction which yields the highest number of $\nu_e$ events is $\nu_e e^-$ elastic scattering in Super-K.  Ironically the copious number of inverse beta interactions makes it hard to isolate this important signal in Super-K.  The addition of gadolinium in Super-K will allow the identification of individual inverse beta events via neutron tagging with high efficiency\,\cite{Beacom:2003nk,Sekiya:2017lgj}.  This will greatly improve the detection prospects of $\nu_{e}$ because gadolinium will allow tagging and event-by-event subtraction of much of the inverse beta background signal. Reference \cite{Laha:2013hva} demonstrated the improvements of this technique for the first time and showed that it is possible to extract the spectral parameters for the CCSN $\nu_{e}$ spectrum to $\sim$20\% precision.  In the future, the lead-based HALO-2 detector \cite{Duba:2008zz,Vaananen:2011bf} and the argon-based DUNE experiment \cite{Acciarri:2015uup,Ankowski:2016lab,Acciarri:2016ooe,Kemp:2017kbm} will detect a large number of CCSN $\nu_e$ events with minimal competing backgrounds and are expected to yield clean measurements of the spectral parameters of CCSN $\nu_e$.

In Ref.\,\cite{Laha:2013hva}, mock $\nu_e$ spectra were generated using analytical and time-averaged expressions for some fiducial CCSN neutrino spectra.  We generalize the approach by generating mock spectra from CCSN simulation data, in which the CCSN neutrino spectrum varies with time\,\cite{Nakamura:2014caa,Nakamura:2016kkl}.  We also explicitly include an additional degree of freedom to the spectrum, the shape parameter $\alpha$, which controls the shape of the emission spectrum [see Eq.~(\ref{eq:maxBolt})].  Reference \cite{Laha:2013hva} included $\alpha$, but it was held fixed at 3, whereas here we allow it to vary in time according to the simulation. The goal of this paper is to explore the usefulness of Super-K, with and without added gadolinium, DUNE, as well as Hyper-K, with and without inverse beta tagging, to extract the spectral parameters of CCSN $\nu_{e}$'s from a time-dependent situation that more directly reflects ongoing simulation efforts. 

In Sec.~\ref{sec:emission}, we discuss neutrino production from a CCSN as well as the simulation we used. In Sec.~\ref{sec:mockData}, we discuss our mock data generation from the simulation data. In Sec.~\ref{sec:chiSquare}, we apply the chi-squared test to the mock data, and we conclude in Sec.~\ref{sec:conclusions}.

\section{Supernova neutrino emission} \label{sec:emission}

\subsection{Neutrino spectrum}

A precise prediction of the CCSN neutrino spectrum depends on a robust understanding of the CCSN microphysics and explosion mechanism.  Recent years have seen a tremendous amount of progress in understanding the CCSN phenomenon. While the problem is not yet fully solved, simulations are starting to show robust explosions \cite{Hix:2016qoa,Maund:2017hvo,Janka:2017vlw,Janka:2017vcp,Burrows:2016ohd,Kuroda:2017trn,OConnor:2015rwy,OConnor:2014sgn,Takiwaki:2016qgc}. Based on our current understanding, the neutrino spectrum can be approximated with reasonable accuracy by a pinched thermal distribution with an associated mean energy. We use a modified normalized Fermi-Dirac distribution of the form \cite{Keil:2002in}
\begin{equation} \label{eq:maxBolt}
\frac{dN_\nu}{dE_\nu}(E_{\nu})=A\bigg(\frac{E_{\nu}}{\langle E_{\nu}\rangle}\bigg)^{\alpha}\text{exp}\bigg[-(\alpha+1)\frac{E_{\nu}}{\langle E_{\nu}\rangle}\bigg],
\end{equation}
where $A=\dfrac{(\alpha+1)^{\alpha+1}}{\langle E_{\nu}\rangle\Gamma(\alpha+1)}$ normalizes the equation to unity, $E_{\nu}$ and $\langle E_{\nu}\rangle$ are the individual neutrino energy and the average neutrino energy both in units of MeV, and $\alpha$ is referred to as the shape parameter because it controls how tightly peaked the spectrum is \cite{Keil:2002in,Tamborra:2012ac}.  There are other ways to fit a CCSN neutrino emission spectrum, but the pinched Fermi-Dirac distribution provides a good match to simulation results \cite{Raffelt:2002tu,Tamborra:2012ac}.

To normalize the spectrum in Eq.\,(\ref{eq:maxBolt}), we need the total energy, $E_{\nu}^{\rm tot}$, which we present in units of $10^{52}$ ergs.  The resulting neutrino fluence per flavor at a distance $d$ from the CCSN is obtained by multiplying Eq.\,(\ref{eq:maxBolt}) by the total particle number, $E_{\nu}^{\rm tot}/\langle E_{\nu}\rangle$, and dividing by $4\pi d^{2}$. We use $d=10$ kpc as the canonical distance from the Galactic CCSN to detector.  The neutrino fluence per flavor at Earth, in units of ${\rm MeV}^{-1} \, {\rm cm}^{-2}$, is
\begin{equation} \label{eq:fluence}
\Phi(E_{\nu})=\frac{1}{4\pi d^{2}}\frac{E_{\nu}^{\rm tot}}{\langle E_{\nu}\rangle}\frac{dN_\nu}{dE_\nu}(E_{\nu}).
\end{equation}
We improve on previous work and use realistic values of $\alpha$, $\langle E_{\nu}\rangle$, and $E_{\nu}^{\rm tot}$ from a CCSN simulation, and thus all three of them vary with time.  As a result, the neutrino incident on Earth is also a function of time.  In Sec.~\ref{sec:simData}, we discuss how the time dependence is taken into account, and we will see later that not accounting for this can cause biases in the final analysis. For a detailed discussion on neutrino production in CCSNe, see Refs.\,\cite{Thompson:2002mw,Tomas:2004gr,Suwa:2008sf,Marek:2008qi,Huedepohl:2009wh}.

\subsection{Simulation data} \label{sec:simData}

We adopt a long-term axisymmetric core-collapse simulation starting from the nonrotating solar metallicity progenitor of Woosley et al.~\cite{Woosley:2002zz} with zero-age main sequence mass of $17 M_\odot$. This progenitor retains its hydrogen envelope at the onset of collapse, making it appropriate as a case study of the most common Type II supernova. The numerical code is the same as that found in Ref.~\cite{Nakamura:2014caa} and updated in Ref.~\cite{Nakamura:2016kkl} specifically for the purpose of following the late-phase evolution, e.g., increased outer spatial boundary. For $\nu_e$ and $\bar{\nu}_e$ transport, the isotropic diffusion source approximation (IDSA) \cite{Liebendoerfer:2007dz} is employed, whereas for heavy-lepton neutrinos a leakage scheme is used. This leakage approximation limits the ability to reliably extract spectral information of the $\nu_x$. Thus, it is assumed that its temperature is given by the temperature of matter at the average $\nu_x$ neutrinosphere, and the shape parameter is kept fixed at $2.3$. In the high-density regime, the equation of state of Lattimer and Swesty \cite{Lattimer:1991nc} with nuclear incompressibility of $K=220$ MeV is used. Explosive nucleosynthesis is followed by a simple network of $13$ alpha-nuclei, and the energy feedback is taken into account. We refer the reader to Refs.~\cite{Nakamura:2014caa,Nakamura:2016kkl} for further details. The simulation lasts for about 7 seconds when the shock reaches the simulation boundary of 100,000 km, roughly corresponding to the bottom of the helium layer. 

\begin{figure}[t]
\centering
\includegraphics[width=\linewidth]{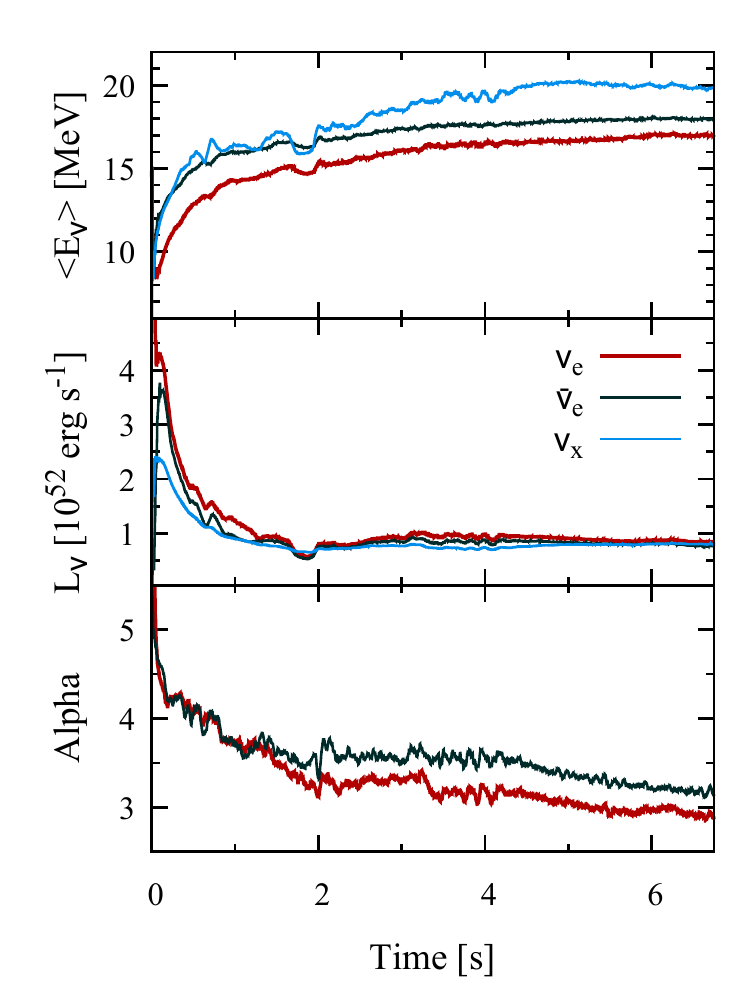}
\caption{Spectral parameters as functions of time for different neutrino types in the simulation used: $\nu_e$ (thick dark red), $\bar{\nu}_e$ (dark green), and $\nu_x$ (light blue).  Here, $\nu_{x}$ represents $\nu_{\mu}$ or $\nu_{\tau}$ or their antiparticles (i.e., they are not added together).  The luminosity of $\nu_{e}$ peaks around $20 \times 10^{52}$ erg (beyond the scale of the figure) at the start of the simulation, due to the neutronization burst during collapse. For $\nu_{x}$, we hold $\alpha$ fixed at 2.3 for the entire time, and it is not shown in the bottom panel.}
\label{fig:simData}
\end{figure}

We first generate spectral parameters for the time-integrated neutrino emission predicted by the simulation. At each time step in the simulation, for each neutrino type, the values for the three spectral parameters $(\langle E_{\nu}\rangle_i, \alpha_i, E_{\nu,i}^{\rm tot})$ are used to compute Eq.~(\ref{eq:fluence})---here, $E_{\nu,i}^{\rm tot}$ becomes the instantaneous luminosity, $L_{\nu}$, multiplied by the length of each time step---giving a neutrino spectrum for that time step, $\Phi_{i}(E_{\nu})$. Then, to get the time-integrated fluence, which we will call $\Phi_{T}(E_{\nu})$, we simply sum, i.e., $\Phi_{T}(E_{\nu})=\sum_{t_{i}}^{t_{f}}\Phi_{i}(E_{\nu})$, from the start ($t_i$) to the finish ($t_f$) of the simulation. This is also how we find the total emitted energy, $E_{\nu,T}^{\rm tot}=\sum_{t_{i}}^{t_{f}}E_{\nu,i}^{\rm tot}$. The parameters $\langle E_{\nu}\rangle_T$ and $\alpha_T$ of $\Phi_{T}(E_{\nu})$ are then computed from the energy moments of the \textit{normalized} (to unity) time-integrated spectrum, $dN_\nu/dE_\nu(E_{\nu})_T$. Therefore, 
\begin{equation} 
\langle {E}_{\nu}\rangle_T=\int_{0}^{\infty} E_{\nu} \frac{dN_\nu}{dE_\nu}(E_{\nu})_T \,d E_{\nu},
\end{equation}
and $\alpha_T$ is defined using the relation \cite{Tamborra:2012ac,Keil:2002in}
\begin{equation} \label{eq:energyMoments}
\frac{\langle E_{\nu}^{k}\rangle_T}{\langle E_{\nu}^{k-1}\rangle_T}=\frac{k+\alpha_T}{1+\alpha_T}\langle E_{\nu}\rangle_T,
\end{equation}
where the energy moments are defined by
\begin{equation} 
\langle E_{\nu}^{k}\rangle_T=\int_{0}^{\infty} E_{\nu}^{k} \frac{dN_\nu}{dE_\nu}(E_{\nu})_T \,d E_{\nu}.
\end{equation}
Using $k=2$ and solving for $\alpha_T$ in Eq.~(\ref{eq:energyMoments}), we find that for $\nu_{e}$, $\alpha_T=2.67$. Higher values of $k$ yield similar results (e.g., $k=3 \rightarrow \alpha_T=2.54$); we will use $k=2$ throughout this paper. The final spectral values are summarized in Table \ref{tab:simDataTable}. In general, the $\alpha_T$ values in the Table are systematically lower than the simulation outputs (bottom panel of Figure \ref{fig:simData}). This is due to the time variation of the spectral parameters; i.e., by summing contributions from different times with different mean energies, luminosities, and shape parameters, the final spectrum is broader than the spectrum at any single snapshot, and this is reflected in the smaller $\alpha_T$. 

For clarity, when a parameter is referenced with subscript $T$, this indicates that we are referring to the time-integrated value, i.e., the value obtained using the time-summed simulation prediction $\Phi_{T}(E_{\nu})$ and the methods outlined in this section. When the subscript is absent, we are simply referring to the parameter itself. 

The last column in Table \ref{tab:simDataTable} shows the total number of particles emitted from the  CCSN  in each flavor (here, $\nu_{x} =  \nu_{\mu} + \bar\nu_{\mu} + \nu_{\tau} + \bar\nu_{\tau}$), found by $E_{\nu,T}^{\rm tot}/\langle E_{\nu}\rangle_T$ after proper unit conversions.

\begin{table}
\centering
\begin{tabular}{ccccc}
\hline \hline $\nu$ type & $\langle E_{\nu}\rangle_T$ [MeV] & $E_{\nu,T}^{\rm tot}$ [$10^{52}$ erg] & $\alpha_T$ & $N_T$ [$10^{57}$] \\
\hline $\nu_{e}$ & 14.1 & 7.70 & 2.67 & 3.41 \\
$\bar{\nu}_{e}$ & 16.3 & 6.44 & 3.28 & 2.47 \\
$\nu_{x}$ & 17.2 & 5.88 & 2.20 & 8.53 \\
\hline \hline
\end{tabular}
\caption{Spectral parameters for the time-integrated spectra of each neutrino type.  Here, $\alpha_T$ is estimated from the second energy moment, and $N_T$ represents the total particle number, which is equal to $E_{\nu,T}^{\rm tot}/\langle E_{\nu}\rangle_T$ after proper unit conversions.}
\label{tab:simDataTable}
\end{table}

\section{Mock data generation} \label{sec:mockData}

We consider three large neutrino detectors: Super-K with and without gadolinium loading, Hyper-K with and without neutron tagging, and DUNE.  We do not explicitly include JUNO because it has worse capabilities compared to Super-K for $\nu_e$ \cite{Laha:2014yua,Fischer:2015oma,Lu:2016ipr}.  Super-K is currently one of the best options for detecting CCSN  neutrinos, with a fiducial mass of 32 kton. Hyper-K, which is in the proposal stage, will be a bigger version of Super-K, with a fiducial mass of 370 kton for CCSN neutrino detection (11.6 times that of SK) \cite{Hyper-Kamiokande:2016dsw}. DUNE is a liquid argon detector with a fiducial mass of 40 kton \cite{Ankowski:2016lab,Acciarri:2016ooe}. We do not include HALO-2, a 1 kton lead-based detector that will detect a clean $\nu_e$ signal but with far fewer event statistics than DUNE \cite{Duba:2008zz,Vaananen:2011bf}.

All neutrino flavors will undergo elastic electron scattering, $\nu e^{-}\rightarrow \nu e^{-}$, although the cross section is larger for electron-type neutrinos due to the additional charged-current contribution. Typically, this interaction produces the largest number of $\nu_e$ events in Super-K and Hyper-K.  The recoiling electron has a kinetic energy that is dependent on the scattering angle $\theta$ as well as the neutrino energy $E_{\nu}$: $\text{cos}\,\theta=\sqrt{T_{e}/(T_{e}+2m_{e})}(E_{\nu}+m_{e})/E_{\nu}$, where $T_{e}$ is the recoil kinetic energy of the electron and $m_{e}$ is the electron mass.  Depending on the scattering angle $\theta$, the range of $T_{e}$ is from 0 to $E_{\nu}^{2}/(m_{e}+2E_{\nu})$.  The detected events are forward peaked and can be used to reduce backgrounds by identifying the elastic scattering events in the forward cone.

The elastic scattering, $\nu e^- \rightarrow \nu e^-$, has the differential cross section\,\cite{Vogel:1989iv},
\begin{eqnarray} \label{eq:dSigdTeScat}
\frac{d\sigma}{dT_{e}}=\frac{G_{F}^{2}m_{e}}{2\pi} && \bigg[(g_{V}+g_{A})^{2}+(g_{V}-g_{A})^{2}\bigg(1-\frac{T_{e}}{E_{\nu}}\bigg)^{2} \nonumber \\
&&+(g_{A}^{2}-g_{V}^{2})\frac{m_{e}T_{e}}{E_{\nu}^{2}}\bigg]\,,
\end{eqnarray}
where $G_{F}$ is the Fermi coupling constant, and $g_{A}=\pm\frac{1}{2}$ and $g_{V}\,=\,2\,\text{sin}^{2}\theta_{W}\pm\frac{1}{2}$ for $\nu_{e}$ and $\nu_{x}$ respectively. For each neutrino's corresponding antineutrino, $g_{A}$ is negated and $g_{V}$ remains the same. 

An additional charged-current interaction of $\nu_e$ in water Cherenkov detectors is $\nu_{e}+{}^{16}\text{O}\rightarrow e^{-}+{}^{16}\text{F}^{*}$. The $^{16}\text{F}^{*}$ itself is undetectable, but the recoil electron is detected via its Cherenkov light. This reaction has a threshold of $15.93\:\:\text{MeV}$, which yields a recoil electron of kinetic energy $T_{e}=E_{\nu}-15.93\:\:\text{MeV}$.  The reaction is very nearly isotropic, and for simplicity we do not consider the anisotropy in our analysis.  The cross section for this interaction, in the energy range 25 MeV $\leq E_{\nu} \leq$ 100 MeV, is\,\cite{Haxton:1987kc} 
\begin{eqnarray} \label{eq:crossSecO}
\sigma(E_{\nu})\approx4.7\times10^{-40}(E_{\nu}^{\nicefrac{1}{4}}-15^{\nicefrac{1}{4}})^{6}\:\:\text{cm}^{2}\,.
\end{eqnarray}
We extrapolate this cross section formula to lower energies and assume that it is close to the true cross section below 25 MeV. The other neutrino interactions with oxygen nuclei have smaller yields and are neglected in this analysis\,\cite{Balasi:2015dba,Langanke:1995he,Nussinov:2000qc}.

The largest number of neutrino interactions are from the inverse beta decay (IBD), $\bar{\nu}_{e}+p\rightarrow e^{+}+n$.  This interaction is nearly isotropic (we again do not consider anisotropy in the analysis) in water Cherenkov detectors. In the current Super-K, this provides the main background for $\nu_{e} e^-$ elastic scattering interaction detection.  The cross section for this reaction is \cite{Vogel:1999zy,Strumia:2003zx}
\begin{equation} \label{eq:crossSecIB}
\sigma(E_{\nu})\approx9.52\times10^{-44}(E_{\nu}-1.3)^{2}\bigg(1-\frac{7E_{\nu}}{m_{p}}\bigg)\:\:\text{cm}^{2},
\end{equation}
where $m_{p}$ is the proton mass. The inverse beta interaction has a threshold of $1.81\,\text{MeV}$, and therefore $T_{e}=E_{\nu}-1.81\:\text{MeV}$.

Recently, the Super-K collaboration has approved the addition of gadolinium to the detector in order to improve various physics analyses\,\cite{Beacom:2003nk}.  Gadolinium has a large capture cross section for neutrons.  A temporal and spatial coincidence of the neutron capture signal on gadolinium and the Cherenkov light from the positron allows for an unambiguous detection of the IBD interaction.  The formidable IBD background for $\nu_{e} e^-$ elastic scattering analysis can therefore be removed.  The tagging efficiency in the Super-K detector is proposed to reach 90\% efficiency, and we use this in our analysis.  We also consider Hyper-K with and without neutron tagging, which may be achieved with gadolinium (the addition of gadolinium has not yet been officially confirmed) or improved photon coverage, allowing improved tagging captures on protons. 

In DUNE, CCSN $\nu_e$ can be detected via the charged-current reaction $\nu_{e}+{}^{40}\text{Ar}\rightarrow e^{-}+{}^{40}\text{K}^{*}$ \cite{Bueno:2003ei,GilBotella:2003sz,Chauhan:2017tgf}. We use the numerical cross section for this interaction from Ref.~\cite{Cocco:2004ac}, which is based on Refs.~\cite{Ormand:1994js} and \cite{Kolbe:2003ys} at low and high energies, respectively.  A more recent cross section can be found in \cite{Chauhan:2017tgf}, which is slightly lower in magnitude than the one we have used.  Importantly, the proposed CAPTAIN experiment aims to measure the neutrino-argon cross section at CCSN energies for the first time\,\cite{McGrew,Whitehead}. We assume that, in DUNE, the $\nu_e$ CC events will be tagged by a photon-coincidence signal\,\cite{Ankowski:2016lab}.  The exact efficiency of this tagging is still undetermined; here, we take it to be a generous 100\%. 

For Super-K, we adopt an electron-detection threshold of 3 MeV\,\cite{Abe:2016waf}.  In Hyper-K, according to the most recent design report, the threshold energy is also expected to be 3 MeV.  This is reduced from the previously expected threshold energy of 10 MeV\,\cite{Abe:2011ts}, which has a dramatic impact on the final results.  In DUNE, we adopt a threshold energy of 5 MeV\,\cite{Ankowski:2016lab}.

\begin{figure}[t]
\centering
\includegraphics[width=\linewidth]{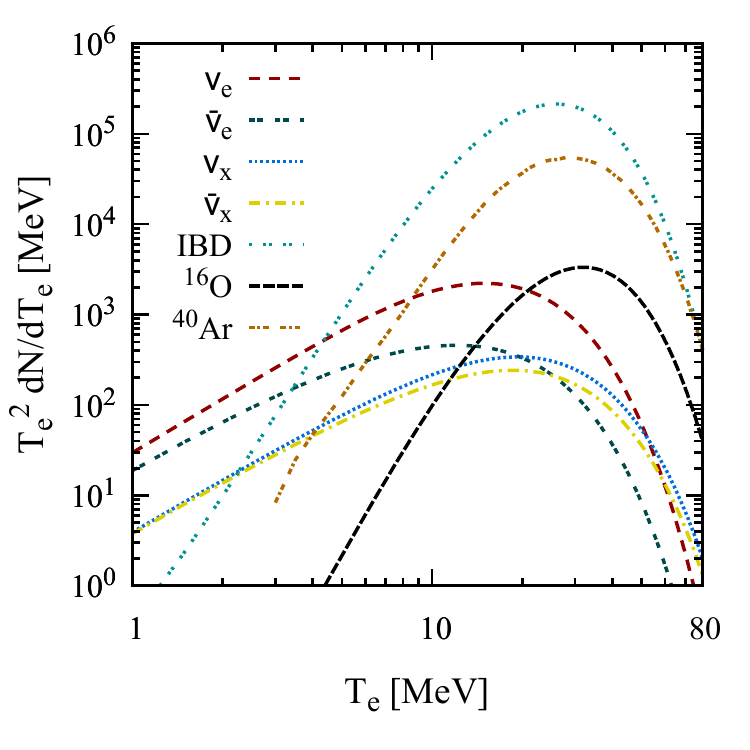}
\caption{Mock detector spectra $T_e^2 \, dN/dT_e$ against lepton kinetic energy $T_e$ for all detection channels considered. For all channels except ${}^{40}$Ar (exclusive to DUNE), parameters from Super-K are used, with gadolinium not included (if included, it would reduce the IBD curve by a factor of 10). Hyper-K's spectra are identically shaped and only multiplied by a factor of 11.6. These spectra are obtained from the whole detector volume; in Sec.~\ref{sec:chiSquare}, we apply an angular cut to the detector data. The $^{40}$Ar spectrum starts at 3 MeV because the numerical cross section we use does not go lower than 3 MeV.}
\label{fig:logLogSpectra}
\end{figure}

The observed event spectrum is given by \cite{Laha:2013hva}
\begin{equation} \label{eq:dNdTeScat}
\frac{dN}{dT_{e}}=N_{t}\int_{E_{min}}^{\infty}dE_{\nu}\Phi(E_{\nu})\frac{d\sigma}{dT_{e}}(E_{\nu}),
\end{equation}
where $N_{t}$ is the number of targets in the detector for a given reaction (e.g., electrons for electron scattering), and $E_{min}$ is the minimum neutrino energy as a function of $T_{e}$, i.e., the minimum required neutrino energy to produce a recoil electron with energy $T_{e}$.  In Fig.~\ref{fig:logLogSpectra}, the mock detector spectra are shown for all detection channels between the three detectors.  In this figure, the entire detector volume was used to obtain the spectra, whereas in Sec.~\ref{sec:chiSquare} we apply an angular cut to the Super-K and Hyper-K data, which optimizes the signal-to-background ratio.  The IBD curve reflects Super-K without gadolinium, which if included would reduce the IBD curve by a factor of 10.

Table \ref{tab:eventCounts} shows the total event count for each detection channel in each detector. Since Hyper-K is bigger than Super-K, it detects many more events and, therefore, offers better statistics in the final analysis, which has a significant impact on the final results.

\section{Analysis of mock data} \label{sec:chiSquare}

\subsection{Full simulation}


In order to infer the measurements of spectral parameters $\langle E_{\nu}\rangle_T$, $E_{\nu,T}^{\rm tot}$, and $\alpha_T$, as well as compare the precision and accuracy of each detector configuration, we construct a $\chi^{2}$ fit test for the $\nu_{e}$ detector spectrum. We use the following chi-squared statistic:
\begin{equation} \label{eq:chiSquare}
\chi^{2}=\sum_{j}\bigg(\frac{O_{j}-T_{j}[\langle E_{\nu}\rangle, E_{\nu}^{\rm tot},\alpha]}{\sigma_{j}}\bigg)^{2}\,,
\end{equation}
where $O_{j}$ is the number of observed signal events in the $j$-th electron recoil energy bin assuming fiducial $\nu_e$ neutrino spectra, and $T_{j}[\langle E_{\nu}\rangle, E_{\nu}^{\rm tot},\alpha]$ is the same for given different values for $\langle E_{\nu}\rangle$, $ E_{\nu}^{\rm tot}$, and $\alpha$. The uncertainty on the observed event numbers in the $j$-th bin is denoted by $\sigma_{j}$, and this includes background event counts that are dependent on the detector setup. 

\begin{table}
\centering
\begin{tabular}{lrrr}
\hline \hline Channel & Super-K & Hyper-K & DUNE \\ 
\hline $\nu_{e}$ scattering & 300 & 3,500 & 260 \\
 $\bar{\nu}_{e}$ scattering & 84 & 970 & 73 \\
 $\nu_{x}$ scattering & 41 & 480 & 36 \\
 $\bar{\nu}_{x}$ scattering & 31 & 370 & 28 \\
 $^{16}\text{O}$ & 110 & 1,300 & $\cdot\cdot\cdot $\\
 IBD & 9,800 & 110,000 & $\cdot\cdot\cdot$ \\
 $^{40}\text{Ar}$ &$ \cdot\cdot\cdot$ & $\cdot\cdot\cdot$ & 2,200 \\
\hline \hline
\end{tabular}
\caption{Approximate total event counts for each channel in each detector. We assume DUNE will be able to subtract the electron scattering events in our analysis, but they are shown here for the sake of comparison.}
\label{tab:eventCounts}
\end{table}

We assume that the $\nu_e + \text{Ar}$ CC events at DUNE will be tagged by a photon-coincidence signal \cite{Ankowski:2016lab} with 100\% efficiency, thus $O_j$ is purely $\nu_e$, and $\sigma_{j}$ will only include the square root of $O_{j}$.  In Super-K and Hyper-K, the $\nu_{e}e^-$ scattering cannot be individually distinguished from $\bar{\nu}_{e}e^-$ and $\nu_{x}e^-$ scatterings.  However, since the spectra of $\bar{\nu}_e$ and $\nu_x$ will be well measured by other distinguishable channels, we assume that their contributions can be statistically subtracted. Hence we only need to fit for the $\nu_e$ spectral parameters, and $O_j$ only depends on the $\nu_e$ parameters \cite{Laha:2013hva}.  The electron elastic scattering of other neutrino flavors, the $^{16}$O interaction and the untagged IBD events, however, contribute towards the backgrounds for $\nu_{e}e^-$ elastic scattering.  We therefore set $\sigma_{j}$ to be the square root of the sum of all the backgrounds per bin, in addition to $O_{j}$.  The IBD events contribute significantly to $\sigma_{j}$, but by adding gadolinium to Super-K or Hyper-K, its contribution will be significantly reduced, resulting in higher overall $\chi^{2}$ values and tighter constraints on the spectral parameters.  For the spectra of $\bar{\nu}_e$ and $\nu_x$, we use the simulation outputs (Fig.~\ref{fig:simData}) and assume they will be measured by IBD and low threshold liquid scintillator detectors, respectively. 

\begin{figure*}[]
\centering
\includegraphics[width=0.65\linewidth]{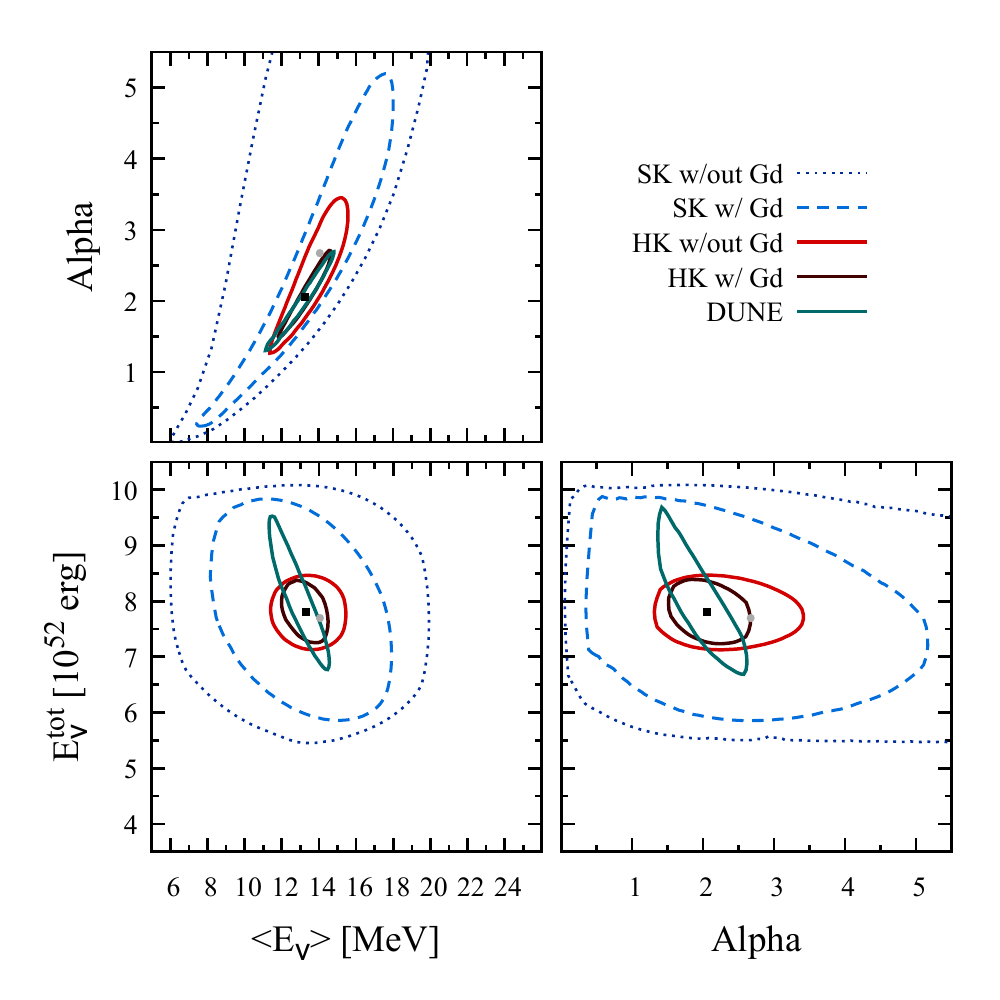}
\caption{90\% C.L. $\Delta\chi^{2}$ contours for the different detector setups. The black square represents the best-fit values from $\chi^{2}_{\rm min}$, $\langle E_{\nu_{e}}\rangle=13.3$ MeV, $E_{\nu_{e}}^{\rm tot}=7.8\times10^{52}$ erg, $\alpha_{\nu_{e}}=2.1$, whereas the gray circle represents values from the simulation (both summarized in Table \ref{tab:bestFitValues}). The offset between them is discussed in the text. Super-K without gadolinium (thin dotted) shows contours that do not fully close in the shape parameter $\alpha$. The addition of gadolinium (dashed) qualitatively changes this. Hyper-K (solid) is a significant improvement on Super-K due to the higher number of events. DUNE's contour (thick solid) is qualitatively different from the other detectors' contours because the $^{40}$Ar reaction probes a different part of the spectrum than the channels in Super-K and Hyper-K. Note that DUNE contours assume 100\% $\nu_e$ CC identification efficiency; in practice the DUNE contours will be larger.}
\label{fig:csTriPlot_All}
\end{figure*}

In order to reduce the Super-K and Hyper-K backgrounds further, we apply an angular cut to the data.  Since electron scattering is a forward-peaked interaction, we can isolate a forward cone that contains the majority of the total elastic scattering events, which will subsequently contain a smaller factor of the isotropic IBD and $^{16}$O backgrounds.  We use a 28$^\circ$ forward cone cut, which includes 68\% of the forward-peaked $\nu_{e}e^-$ elastic scattering events\,\cite{Abe:2010hy}, while only including $\sim 6\%$ of the IBD and $^{16}$O events. This angle changes depending on the average $\nu_{e}$ energy; as $\langle E_{\nu_{e}}\rangle$ increases, the events become more tightly peaked, and so the forward cone angle containing 68\% of the events decreases. In our analysis of the $^{16}$O channel, we take the complementary reverse cone as an angular cut in order to reduce the amount of scattering events, which now act as backgrounds. In Fig.~\ref{fig:logLogSpectra} and Table \ref{tab:eventCounts}, no angular cuts have been applied yet, but we apply these cuts to the data used in Eq.~(\ref{eq:chiSquare}).

To maintain Gaussian statistics, we set energy bin boundaries such that there are at least 10 events in each energy bin.  The energy bin size cannot be smaller than the energy resolution of the detector, which for the detectors we consider is typically about 10\% (i.e., the resolution is 2 MeV at 20 MeV). Using these criteria, we determine binning schemes that work well for each detection channel in each detector.

For each detection channel we iterated through a range of values for $\langle E_{\nu}\rangle$, $E_{\nu}^{\rm tot}$, and $\alpha$, yielding $\chi^{2}$ values as a function of those three parameters. The domain of each parameter is chosen as follows: $0 < \langle E_{\nu}\rangle/ {\rm MeV} < 25$, $0 < E_{\nu}^{\rm tot}/ 10^{52} \, {\rm erg} < 11$, and $0 < \alpha < 6$. We iterate through all three parameters and interpolate the data to obtain $\chi^{2}$ as a function of the parameters. To infer the set of spectral parameter values that most closely reproduce the detector signal, we find the set of parameter values that minimizes the $\chi^{2}$. We designate this as the best-fit, and the associated chi-squared value as $\chi^{2}_{\rm min}$.

\begin{table}
\centering
\begin{tabular}{lcrcrc}
\hline \hline Parameter & SK & SK+Gd & HK & HK+Gd & DUNE \\ \hline
$\langle E_{\nu_e}\rangle$ 	& $\pm50$\% & $\pm40$\% & $\pm15$\% & $\pm10$\% & $\pm10$\% \\
$E_{\nu_e}^{\rm tot}$ 		& $\pm30$\% & $\pm20$\% & $\pm10$\% & $\pm7$\% & $\pm20$\% \\
$\alpha_{\nu_e}$ 			& N/A 	      & $\pm110$\% & $\pm50$\% & $\pm30$\% & $\pm30$\% \\
\hline \hline
\end{tabular}
\caption{Symmetrized fractional uncertainties (90\% C.L.) on measured $\nu_e$ spectral parameters, all as percentages of the best-fit values. Uncertainties are estimated from the projections of the contours of Fig.~\ref{fig:csTriPlot_All}. ``N/A'' indicates the contours do not fully close in our explored parameter range.}
\label{tab:errors}
\end{table}

\begin{figure*}[t]
\centering
\includegraphics[width=0.65\linewidth]{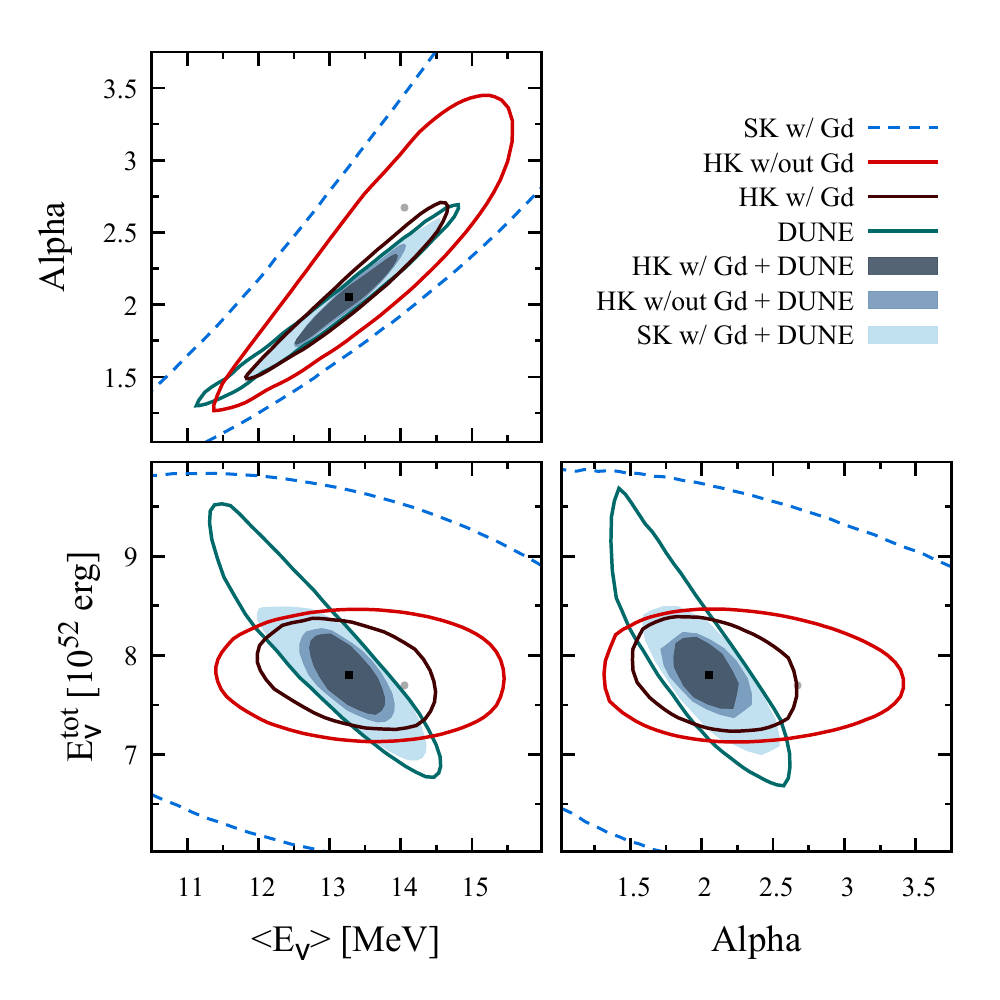}
\caption{Same as Fig.~\ref{fig:csTriPlot_All} but focusing only on Hyper-K and DUNE. Shaded regions show joint 90\% C.L.~regions combining DUNE and Super-K with gadolinium (light blue), DUNE and Hyper-K (blue), and DUNE and Hyper-K with gadolinium (dark blue). Note the different axis scales to Fig.~\ref{fig:csTriPlot_All}.}
\label{fig:csTriPlot_HK}
\end{figure*}

Since we have three parameters of interest, any set of parameters that gives $\Delta\chi^{2} = \chi^{2}-\chi_{\rm min}^{2} < 6.25$ is acceptable within a 90\% confidence level (C.L.). Based on this criterion, we derive contours containing the allowed sets of values. In Fig.~\ref{fig:csTriPlot_All}, we show the 90\% C.L.~contours projected onto two-dimensional planes for all three detectors. In each panel, the parameter that is not shown is integrated over to obtain the contour projection. Take the bottom left panel, for instance. For a fixed value of $\alpha$, the contours for each detector are similar in shape but smaller in size. As $\alpha$ varies, the centers of each contour shifts. The integrated contours shown in Fig.~\ref{fig:csTriPlot_All} represent the area swept out by the fixed-$\alpha$ contours as $\alpha$ is varied over its domain.  Super-K's contours (without gadolinium) do not close on the large $\alpha$ end (top or bottom right panel of Fig.~\ref{fig:csTriPlot_All}) at our chosen 90\% C.L. The contours for Super-K without Gd are also not closed on the low-$\alpha$ end, but we assume that $\alpha\geq0$ for simplicity.  This example clearly shows the comparative improvement of Super-K with Gd compared to Super-K where Gd is not added.  Hyper-K and DUNE, as well as Super-K with gadolinium, have completely closed contours in all parameter directions at 90\% C.L. The improvements are quantified in Table \ref{tab:errors} where we show the fractional 90\% C.L.~uncertainties on the spectral parameters. Although the contours show some asymmetries, we show the mean of the positive and negative uncertainties for simplicity. 

As expected, Hyper-K is a significant improvement on Super-K, regardless of whether or not gadolinium is present.  DUNE is noticeably different in shape than the water Cherenkov detectors. This is because the argon channel probes a different part of the spectrum than the other channels. Since the argon reaction has a threshold of $\sim5.9$ MeV, and DUNE has a detector threshold of 5 MeV, we only see events from neutrinos whose energies are higher than 10.9 MeV. In Super-K and Hyper-K, both the electron scattering and the oxygen channels are present.  Since neutrinos can scatter off electrons at effectively any energy (as long as it is greater than the binding energy of the electron, which is much smaller than the neutrino energy), the only limitation to what we see is the detector threshold, and so the scattering channel probes neutrinos of energies $\gtrsim$ 3 MeV, the threshold for both Super-K and Hyper-K.  The oxygen channel (reaction threshold of $\sim16$ MeV) probes $\nu_e$'s of energies greater than 19 MeV, but combined with electron scattering, we can still see the lower-energy neutrinos in the detector as a whole.

The shape of the contours can be understood by the dependence of the interactions involved on the neutrino emission properties. In water Cherenkov detectors, the elastic scattering and oxygen interaction act in complementary ways. The average energy and the total energy of the $\nu_e$ spectrum are more strongly constrained by the oxygen interaction and elastic scattering, respectively. The number of $\nu_e e$ elastic scattering events is proportional to the product of the number of incident $\nu_e$, which can be approximated by $E^{\rm tot}_{\nu_e, T}/\langle E_{\nu_e} \rangle_T$, and the spectrum averaged interaction cross section. Since the $\nu_e e$ elastic scattering cross section is proportional to the neutrino energy in the leading order [see Eq.~\ref{eq:dSigdTeScat}], the number of $\nu_e e$ elastic scattering events is independent of $\langle E_{\nu_e} \rangle_T$ to the leading order. Thus the $\nu_e e$ elastic scattering events provide only weak constraint on the $\langle E_{\nu_e} \rangle_T$, while providing a good handle on $E^{\rm tot}_{\nu_e, T}$. On the other hand, the $\nu_e \, ^{16}$O and $\nu_e \, ^{40}$Ar interactions depend on higher powers of $\langle E_{\nu_e} \rangle_T$ and thus produce a stronger constraint on the mean energy.  Although similarly sized, the constraint on the $\langle E_{\nu_e} \rangle_T$ is stronger from DUNE when compared to that of Super-K with Gd.  This additional power arises from the higher number of $\nu_e \, ^{40}$Ar interactions.  For all the interactions, the total number of events in the detector is directly proportional to the $E^{\rm tot}_{\nu_e, T}$, and thus the detector with the highest number of events, Hyper-K, produces the strongest constraint on this parameter.

If  CCSN  neutrinos are detected in multiple detectors, the $\chi^{2}$ of the different detectors can be combined to get a tighter net contour. The black square in Fig.~\ref{fig:csTriPlot_All} represents the best-fit parameter values found by combining the $\chi^{2}$ of Super-K with gadolinium, Hyper-K with gadolinium, and DUNE, which is the most ideal future situation for  CCSN  $\nu_{e}$ detection. The resulting best fit values for the neutrino spectral parameters are found in Table \ref{tab:bestFitValues}.

Figure \ref{fig:csTriPlot_HK} zooms in on just the contours for Hyper-K and DUNE for better detail. These contours provide a fairly tight constraint on the possible values for all the $\nu_e$ spectral parameters.  Due to the larger size of the Hyper-K detector and the high efficiency neutron tagging (if present), the contours obtained from Hyper-K are smaller than or comparable to that of DUNE.  Thus, our analysis shows that Hyper-K can be the world-leading detector to constrain the  CCSN  $\nu_e$ spectrum. Also in this figure are shown the joint 90\% C.L.~regions obtained by combining detectors: DUNE and Super-K with gadolinium, DUNE and Hyper-K, and DUNE and Hyper-K with gadolinium. These are computed by the criteria $\Delta \chi^2 < 6.25$ with respect to the minimum in the combined $\chi^2$ plane. As expected, combining detectors is advantageous and reduces the parameter region. Most notably, the elongated DUNE contour is reduced significantly even by the addition of Super-K with gadolinium, highlighting the qualitative advantages of combining detectors with different detection channels and backgrounds. 

\begin{table}
\centering
\begin{tabular}{lcccc}
\hline \hline
& & $\langle E_{\nu_e}\rangle$ [MeV] & $E_{\nu_e}^{tot}$ [$10^{52}$ erg] & $\alpha_{\nu_e}$ \\ \hline 
No mix & Black square & 13.3 & 7.8 & 2.1 \\ 
& Gray circle & 14.1 & 7.7 & 2.7 \\ \hline 
IH & Black square & 14.4 & 6.6 & 1.3 \\ 
& Gray circle & 15.9 & 6.4 & 2.0 \\ \hline 
NH & Black square & 16.8 & 5.9 & 1.8 \\ 
& Gray circle & 17.2 & 5.9 & 2.0 \\ \hline \hline
\end{tabular}
\caption{Best-fit (black square) and time-integrated values (gray circle) for the three different mixing scenarios we consider. The offsets between best fit and time integrated arises due to time variation of the emission, as discussed in the text. For the same reason, the shape parameters are lower than those predicted by the simulation, shown in Fig.~\ref{fig:simData}.}
\label{tab:bestFitValues}
\end{table}

In Fig.~\ref{fig:csTriPlot_HK}, the black square is the same point as in Fig.~\ref{fig:csTriPlot_All}, and the gray circle  represents the spectral values in Table \ref{tab:simDataTable}, i.e., found from the simulation data itself (see also Table \ref{tab:bestFitValues} for values). There is a $\lesssim$ 25\% offset in these values from the $\chi^{2}_{\rm min}$ values.  The threshold energies affect the size of the chi-squared contours but not generally the locations of their minima. Each binning scheme began at the threshold energy for each detector (SK - 3 MeV; HK - 3 MeV; DUNE - 5 MeV), since with actual detector data we would not be able to use anything below the threshold. This means that the results will match the detector spectrum very well above the threshold, but not necessarily below it. Performing a chi-squared test in a no-threshold scenario reveals that removing the detector threshold makes the allowed region slightly smaller, but it does not appreciably change the location of the minimum, meaning that the threshold does not contribute to the offset between the black squares and gray circles.  The origin of the offset is discussed in Sec.~\ref{subsec:timeSplit}

\subsection{Flavor oscillations}

\begin{figure}[t]
\centering
\includegraphics[width=\linewidth]{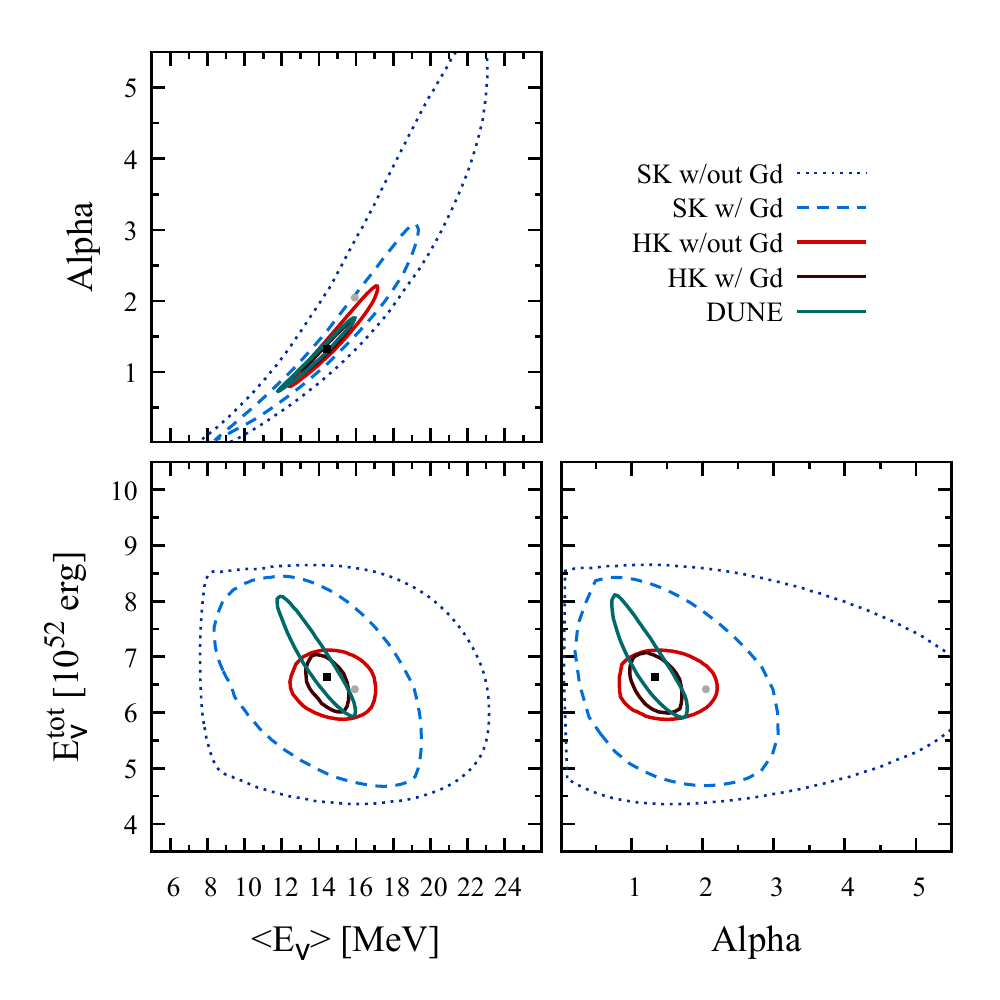}
\caption{Same as Fig.~\ref{fig:csTriPlot_All} but after an IH mixing scheme has been applied. Now, most notably, Super-K without gadolinium has closed contours in all parameter directions (assuming $\alpha\geq0$).}
\label{fig:csTriPlot_All_IH}
\end{figure}

We have not included neutrino flavor oscillations so far in this analysis. We can implement a simplified model of oscillations to get an idea for what effects oscillations might have on the final detection \cite{GilBotella:2003sz}. Here, we use a model based on MSW resonances using both an inverted mass hierarchy (IH) and a normal mass hierarchy (NH)\,\cite{Dighe:2007ks}.  We do not consider collective oscillations as the effect of it on neutrino flavor mixing in  CCSN  is still a matter of debate\,\cite{Chakraborty:2015tfa,Dasgupta:2016dbv,Capozzi:2016oyk,Dasgupta:2015iia,Cherry:2012zw,Duan:2010bf}.  In the MSW scheme, the mixed fluences are found using the following equations\,\cite{Dighe:2007ks}:
\begin{eqnarray}
\Phi_{\nu_e}&=&p\Phi_{\nu_e}^0+(1-p)\Phi_{\nu_x}^0, \\
\Phi_{\bar{\nu}_{e}}&=&\bar{p}\Phi_{\bar{\nu}_{e}}^0+(1-\bar{p})\Phi_{\nu_x}^0, \\
\Phi_{\nu_e}+\Phi_{\bar{\nu}_{e}}+4\Phi_{\nu_{x}}&=&\Phi_{\nu_e}^0+\Phi_{\bar{\nu}_{e}}^0+4\Phi_{\nu_x}^0,  \label{eq:fluenceNuX}
\end{eqnarray}
where $\Phi_{\nu_x}$ represents the fluence of \textit{one} of the four flavors of $\nu_x$, hence the factor of 4 when writing the total fluence in Eq.~(\ref{eq:fluenceNuX}), which can be rearranged to solve for $\Phi_{\nu_x}$. For IH mixing, the survival probabilities are $p=\text{sin}^2\theta_\odot \approx0.3$ and $\bar{p}=0$, where $\theta_\odot$ is the solar neutrino mixing angle. For NH, $p=0$ and $\bar{p}=\text{cos}^2\theta_\odot \approx 0.7$.

\begin{figure}[t]
\centering
\includegraphics[width=\linewidth]{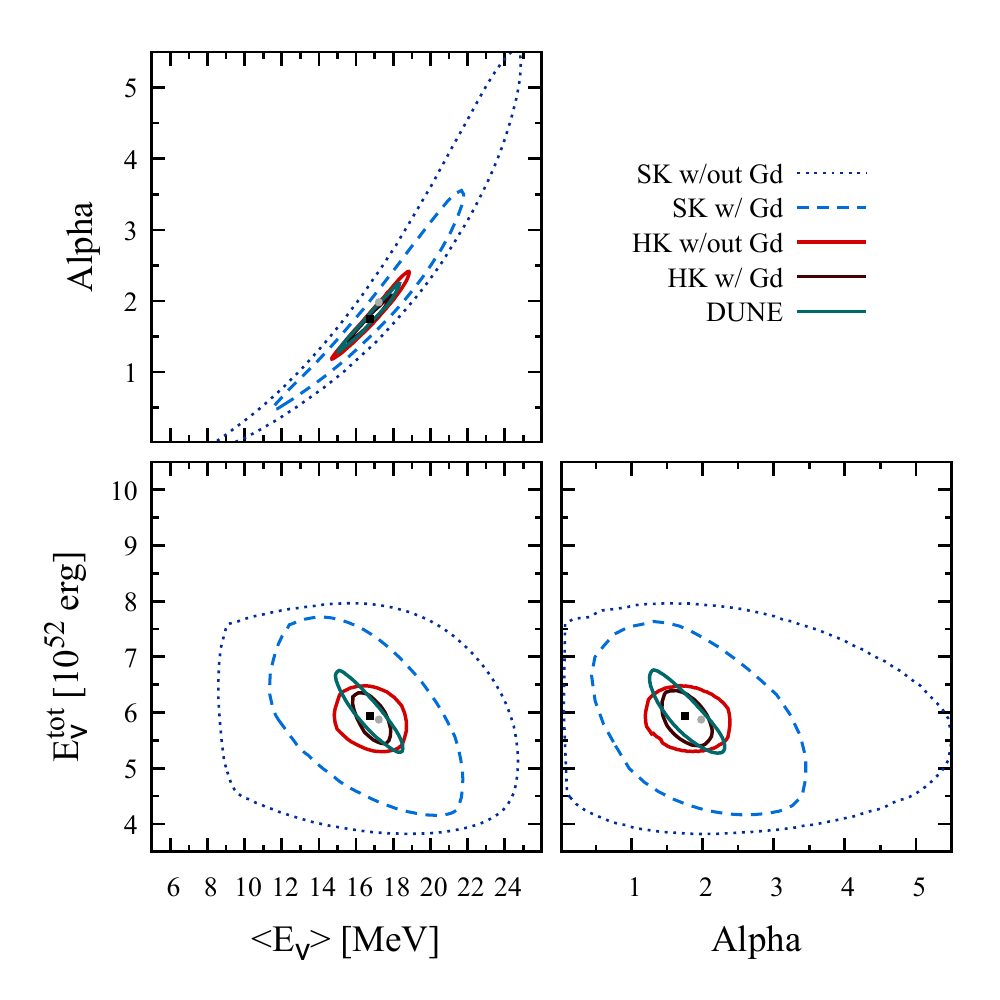}
\caption{Same as Fig.~\ref{fig:csTriPlot_All} but with NH mixing. As with IH, Super-K without gadolinium has closed contours in all parameter directions (assuming $\alpha\geq0$).}
\label{fig:csTriPlot_All_NH}
\end{figure}

The results for IH mixing are shown in Fig.~\ref{fig:csTriPlot_All_IH}. The corresponding best-fit values (black squares) and time-integrated values (gray circles) are found in Table \ref{tab:bestFitValues}. The most notable difference from Fig.~\ref{fig:csTriPlot_All} is that now Super-K without gadolinium has closed contours in all parameter directions.  This is due to the mixing of higher-energy neutrinos that were originally $\nu_x$ towards the detected $\nu_e$, which increases the events.  Capture on oxygen, which has a high energy threshold, introduces qualitative improvements as the neutrino average energy increases above $\sim 15$ MeV.  The contours for NH mixing are shown in Fig.~\ref{fig:csTriPlot_All_NH}. The black square and gray circle values are again found in Table \ref{tab:bestFitValues}. The main difference from IH mixing, other than shifted parameter values, is that the offset between best-fit and time-integrated values is smaller.

\subsection{Analysis in different time epochs}\label{subsec:timeSplit}


In Figs.~\ref{fig:csTriPlot_All}--\ref{fig:csTriPlot_All_NH} and Table \ref{tab:bestFitValues}, we see that there is an offset between the $\chi^{2}_{\rm min}$ parameter values (black square) and the values obtained from spectral analysis of the time-integrated spectra (gray circle), i.e., the values in Table \ref{tab:simDataTable}. This is due primarily to the composite nature of the net emission described in the second paragraph of Sec.~\ref{sec:simData}. Because the net emission is a combination of individual thermal spectra at each time step in the simulation, each with different sets of spectral parameter values, the summed result is no longer a neat thermal spectrum and it is not well-defined by one set of the three spectral parameters we use nor is the detector spectrum it produces.  This is particularly reflected in the shape parameter, which necessarily is reduced to accommodate the broader time-summed spectra. In our analysis, we attempt to fit the detector spectrum using one set of parameter values, so inevitably we will not get a perfect result.

In Super-K, this offset is not a problem because it is well within the precision of the detector. However, with the construction of Hyper-K and DUNE, the offset will become more important. One possibility to minimize the offset is to split the detection signal into smaller time epochs. Within smaller time splices, the spectral parameters do not vary over as large of a range as they do during the whole emission time (see Fig.~\ref{fig:simData}), and so the net spectrum should be closer to an analytical thermal spectrum with one set of parameters.

\begin{figure}[t]
\centering
\includegraphics[width=\linewidth]{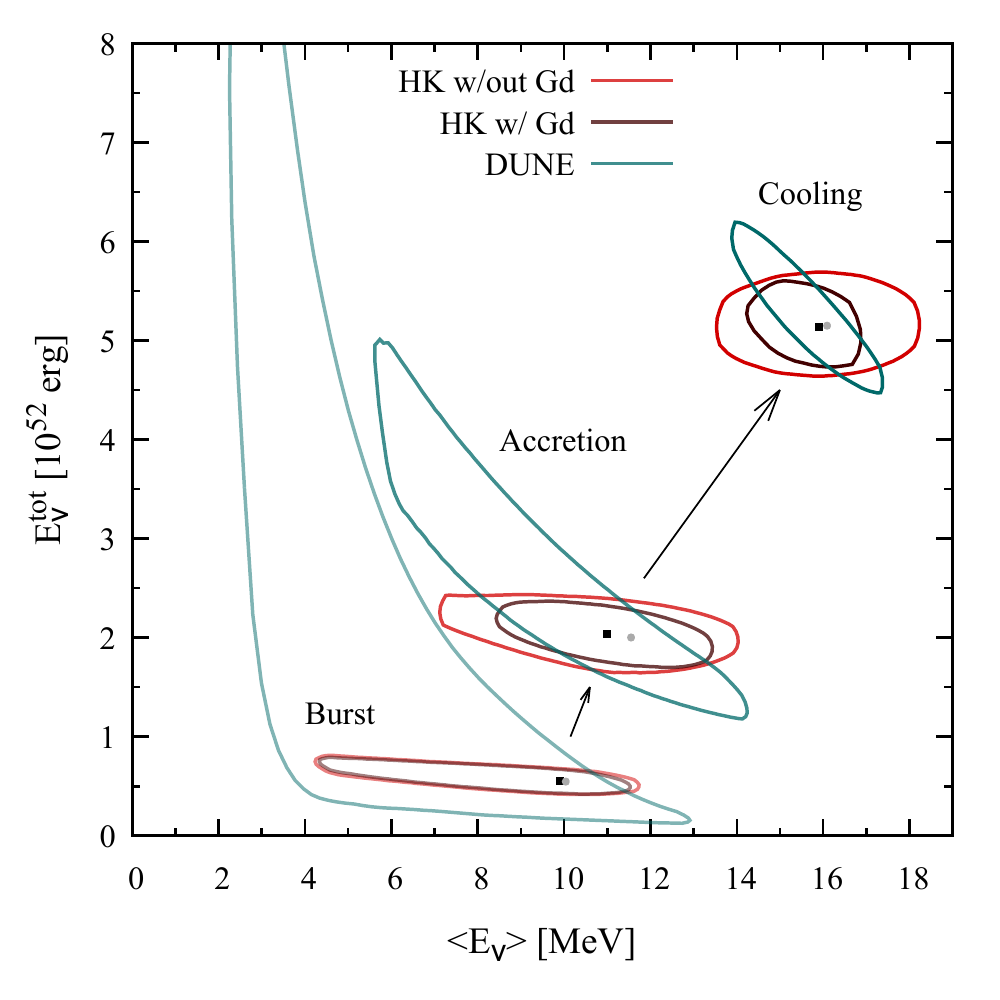}
\caption{$E_{\nu_e}^{\rm tot}$ vs. $\langle E_{\nu_e}\rangle$ contours (i.e., bottom left panel of Fig.~\ref{fig:csTriPlot_All}) for the three time epochs of the supernova burst.}
\label{fig:csPhasesEtotVsEavg}
\end{figure}

We test this hypothesis by splitting the simulation data into three time epochs, namely the neutronization burst, accretion phase, and cooling phase. By looking at Fig.~\ref{fig:simData}, we loosely determine the phase times to be the following: neutronization burst, start -- 60 ms; accretion phase, 60 ms -- 1 s; cooling phase, 1 s -- end. The $E_{\nu_e}^{\rm tot}$ vs. $\langle E_{\nu_e}\rangle$ contours for the three phases are shown in Fig.~\ref{fig:csPhasesEtotVsEavg}. Since each time epoch has lower event statistics than the full simulation, especially the first two phases, the contours are larger in size than the full simulation contours in Fig.~\ref{fig:csTriPlot_All}. However, as expected, in all cases, the $\chi_{\rm min}^{2}$ values (black square) are closer to the simulation values (gray circle), which are shown in Table \ref{tab:phasesData}. Therefore, as we split the supernova signal into time epochs, we trade precision for accuracy, and in a real supernova event, the time splices can, depending on the detection statistics, be adjusted based on desired information, precision, and accuracy. For example, we find that Hyper-K will produce stringent constraints on $\nu_e$ spectral parameters for all three major phases of the supernova.

\begin{table}
\centering
\begin{tabular}{cccc}
\hline \hline $\nu_{e}$ & $\langle E_{\nu}\rangle_T$ [MeV] & $E_{\nu,T}^{\rm tot}$ [$10^{52}$ erg] & $\alpha_T$ \\
\hline Burst & 10.0 & 0.55 & 5.07 \\
Accretion & 11.6 & 2.00 & 3.50 \\
Cooling & 16.1 & 5.15 & 3.12 \\
\hline \hline
\end{tabular}
\caption{Same as Table \ref{tab:simDataTable} but for the three time epochs: neutronization burst, accretion, and cooling phases.}
\label{tab:phasesData}
\end{table}

\section{Conclusions and Discussions} \label{sec:conclusions}

The only way to completely understand a CCSN and to robustly constrain or discover new physics from its neutrino signal is to understand the full flavor emission of  CCSN  neutrinos.  Also, neutrinos escape the  CCSN  with $\sim$99\% of the energy budget, and thus it is extremely important to detect all flavors of neutrinos to account for the full energy emission.  There are well-established ways to robustly detect  CCSN  $\bar{\nu}_e$ and $\nu_x$.  Besides detection via lead or liquid argon detectors,  CCSN  $\nu_e$ has received little attention\,\cite{Laha:2013hva}.

We follow the ideas first presented in Ref.\,\cite{Laha:2013hva} about detecting CCSN  $\nu_e$ in gadolinium-loaded water Cherenkov detectors and go significantly beyond it by using a neutrino spectrum generated from a  CCSN  simulation and by introducing an additional spectral shape parameter $\alpha$.  The use of the forward angle cut and the removal of individual inverse beta events with gadolinium allow for an efficient detection of  CCSN  $\nu_e$, and we have confirmed that the mean energy and total luminosity of a  CCSN  $\nu_e$ emission can be determined with reasonable accuracy of some 10--20\%.  We have now quantitatively shown how the chi-squared contours are affected when the shape parameter $\alpha$ is included as a free parameter. Comparing our results to Ref.~\cite{Laha:2013hva}, which holds $\alpha$ fixed at 3, we see that Super-K is not nearly as precise when $\alpha$ is allowed to vary.  Hyper-K and DUNE offer promisingly precise results even in the face of varying $\alpha$, even more precise than Super-K with fixed $\alpha$ in Ref.~\cite{Laha:2013hva}. Hyper-K and DUNE, therefore, will both offer a clear and immense improvement over Super-K in Galactic  CCSN  neutrino analysis.

Further studies are also needed to explore how uncertainties on $\bar{\nu}_e$ and $\nu_x$ parameters will affect the determination of $\nu$ parameters.  We have also included the effect of MSW transitions in a  CCSN  and analyzed the impact on the reconstruction of the  CCSN  $\nu_e$ spectrum (Figs.~\ref{fig:csTriPlot_All_IH} and \ref{fig:csTriPlot_All_NH}).  

In summary, we show that the three-parameter analysis of CCSN $\nu_e$ that is necessary from a CCSN perspective yields robust results and has considerable promise with near future neutrino detectors. The prospects are very high for the next Galactic CCSN to yield multimessenger signal detections spanning neutrinos, electromagnetic, to gravitational waves \cite{Nakamura:2016kkl}, with potentially extremely rich rewards. We encourage the community to study this new method to study  CCSN  $\nu_e$ in detail and further optimize its tools so that we can learn as much as possible from a future Galactic CCSN.

\begin{acknowledgments}

We thank Ko Nakamura for sharing core-collapse simulation outputs, useful discussions, and careful comments on the manuscript. We thank John Beacom, James Kneller, Kate Scholberg, and Mark Vagins for careful reading of the manuscript and insightful suggestions. R.L.~thanks KIPAC for support.  R.L.~is supported by German Research  Foundation  (DFG)  under  Grant  Nos.  EXC-1098, KO 4820/1-1, FOR 2239, and from the European Research Council (ERC) under the European Union's Horizon 2020 research and innovation programme  (Grant  Agreement  No.  637506,  ``$\nu$Directions") awarded to Joachim Kopp.   S.H.~is supported by the U.S.~Department of Energy under Award No.~de-sc0018327.

\end{acknowledgments}

\bibliography{ms}

\end{document}